# Protecting the Texas power grid from tropical cyclones: Increasing resilience by protecting critical lines

Julian Stürmer[1,2], Anton Plietzsch[1], Thomas Vogt[1], Frank Hellmann[1], Jürgen Kurths[1,3,4], Christian Otto[1,*], Katja Frieler, Mehrnaz Anvari[1,*]

[1]Potsdam Institute for Climate Impact Research, Telegrafenberg A56, 14473 Potsdam, Germany

[2]Institute for Theoretical Physics, TU Berlin, 10623 Germany

[3]Institute of Physics and Astronomy, University of Potsdam, 14476 Potsdam, Germany

[4]Institute of Physics, Humboldt Universität zu Berlin, 12489 Berlin, Germany

*Corresponding author(s): anvari@pik-potsdam.de; christian.otto@pik-potsdam.de

**Abstract**

The Texan electric network in the Gulf Coast of the United States is frequently hit by Tropical Cyclones (TC) causing widespread power outages, a risk that is expected to substantially increase under global warming. Here, we introduce a new approach of combining a probabilistic line fragility model with a network model of the Texas grid to simulate the temporal evolution of wind-induced failures of transmission lines and the resulting cascading power outages from seven major historical hurricanes. The approach allows reproducing observed supply failures. In addition, compared to a static approach, it provides a significant advantage in identifying critical lines whose failure can trigger large supply shortages. We show that protecting only 1% of total lines can reduce the likelihood of the most destructive type of outages by a factor of between 5 and 20. The proposed modelling approach could represent a tool so far missing to effectively strengthen the power grids against future hurricane risks even under limited knowledge.

# Introduction

Modern societies depend heavily on reliable access to electricity. Power outages have the potential to disrupt transportation and telecommunication networks, heating and health systems, the cooling chain underpinning food delivery and more[1–3]. Depending on the cause of power outages and the amount of physical damages to infrastructures, the recovery of the electric network, and the social infrastructures dependent on it, often takes days or even months[4]. Such outages are often driven by extreme weather events. In Norway $90\%$ of all overhead line failures are caused by extreme weather which involves strong winds, icing and lightning strikes[5]. In February 2021 a winter storm in Texas led to outages that in turn caused a breakdown of the gas supply and thus the heating sector[6–8]. Impacts are particularly devastating when it comes to tropical cyclones. In the summer months, the Gulf Coast and the East Coast of the United States are frequently hit by tropical cyclones (TC) that entail widespread outages and costs of billions of dollars. For example, hurricane Ike hitting southeast Texas on September 13, 2008 destroyed around 100 towers holding high voltage transmission lines and cut off electric power for between 2.8 and 4.5 million customers for weeks to months[9, 10]. On August 29, 2021 hurricane Ida made landfall in Louisiana, and destroyed major transmission lines delivering power into New Orleans, causing more than a million customers to lose power[11].

Resilience against line failures in power grids is usually discussed in terms of the N-1 (rarely also N-2) security of the system, that is, the ability of the system to stay fully functional upon the failure of one or two elements[12]. When a line fails, the power flow automatically reroutes through the intact grid. To avoid overloads in the rerouting, relevant lines are intentionally taken out of the grid. This secondary failures of lines can trigger a cascade[13–19] of additional failures. N-1 security asserts that single line failures do not trigger such cascades. Significant secondary failures do occur in larger events and were, e.g., observed in response to the software error leading to the U.S.-Canadian blackout on August 14th, 2003[20]. They are typically also induced by the widespread primary damages and line failures caused by TCs.

The N-1 approach to system resilience does not scale to extreme weather events. The tens or even hundreds of primary failures during events such as hurricanes can not be fully mitigated by an electric network, because N-100 security is not realistic to achieve. N-1 security is typically studied by simulating the reaction of the system to every possible failure scenario. As the number of possible failure scenarios scales exponentially with the number of failures, it is computationally infeasible to consider all possible such scenarios in larger events. Initialising failure cascade models designed for N-1 studies with many initial failures is challenging.

Here, we present an approach that solves these issues by temporally resolving the potential damages induced by hurricanes and a stepwise application of a failure cascade model. This approach particularly allows us to identify critical power lines whose protection could most effectively reduce the risk of severe widespread power outages. Although the frequency of severe hurricanes is expected to increase[12–14], such an approach does not exist so far.

# Main text

Our approach explicitly models the dynamical interplay of an extreme wind event with the power grid. It temporally resolves both, the primary wind damages, and the cascades and secondary failures that result from them. We will use this approach to study the impact of massive TCs on the Texan power grid. Strong hurricanes, such as Harvey that made landfall on Texas and Louisiana in August 2017, can destroy more than hundreds transmission lines in an electric grid (see Fig. 1(a)). These lines do not collapse simultaneously, but over the hours or days the TC passage takes. Making use of the chronological order of the line destructions, we divide each overall TC scenario into a sequence of 5 minute long scenarios. In most of these individual steps, only one line fails. We then solve individual scenarios by representing the Texan transmission network in a DC power flow approximation with conservative load balancing assumptions (see Methods and Supplementary Methods 3 and 4). This approach accounts for the 'path dependency' of the solution: Everytime a line collapses, secondary failures can occur, but also control mechanisms are immediately activated and try to bring back the energy balance to the system and, consequently, mitigate the effect of the failure (see Supplementary Methods 4). Later primary damages along the TC track then meet a partially destroyed, rebalanced grid. Thus, the effect of later failures can be more or, even, less intense. It is the resilience of these intermediate, partially destroyed states that ultimately decides whether the impact of the TC is amplified by secondary failures.

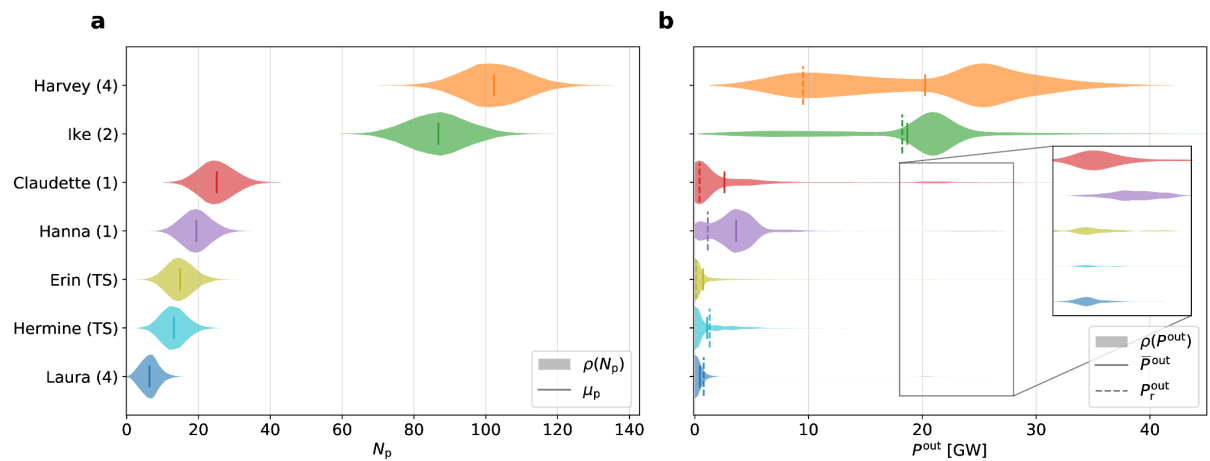

**Fig. 1: Probability distributions of primary line failures and final power outages** (a) Probability distribution of the total number of wind-induced line failures $N_p$ as generated by the probabilistic line fragility model for each of the seven recent hurricanes hitting Texas (category in brackets behind the name). TCs are sorted according to the means of the distributions which are indicated as solid vertical lines. (b) Probability distribution of the associated total power outage $P^{out}$ after TC passage. The inset highlights large cascading failures that can also occur for the weaker hurricanes. The dashed vertical lines indicate the reported power outages listed in the Supplementary Table 1 and the solid vertical lines represent the means. See Methods section for the model parameters used in the simulations.

Unfortunately, neither detailed information about the topology of the exposed power grid nor about the exact power lines destroyed by the considered TC is publically accessible. So here, we use a synthetic model of the Texan grid introduced by Bircheld et al[21] (see Supplementary Fig. 2 as well as Methods).

To represent the TCs impact on the energy supply we combine this grid model with a probabilistic line destruction model (see Methods) forced by modelled historical wind fields from seven different TCs (see Supplementary Supplementary Methods 2). The probabilistic model provides the probability of line failure in terms of wind speeds and allows to generate a large sample of temporally resolved realisations of line failure maps. In the default setting considered here we assume a homogeneous base failure rate for all transmission lines. This is our main adjustable parameter and is tuned to reproduce observed power outages (see Fig. 1(b) and Supplementary Methods 5). The TCs are selected to cover several different types of trajectories and intensities and particularly include storms that continue to move westward after landfall and affect the southern and western parts of Texas such as Hurricane Claudette, Tropical Storm Erin, and Hurricane Hanna, contrary to most hurricanes that are steered northward by the Coriolis effect before western parts of Texas are reached[22].

## Core result

While the number of primary line failures follows a Poisson binomial distribution, the derived distribution of outages is heavily multimodal for all storm tracks with the potential of large $20\,GW$ to $30\,GW$ outages (see Fig. 1(b)). These large damages turn out to not accumulate gradually over the course of the hurricane but occur suddenly in one or few time steps (see Fig. 2(b)). This sudden increase in outages is induced by cascading line failures taking the Houston and a weakly connected North-Western section of the grid offline (see Fig. 2(d) and Fig. 3).

Figure 3 shows what damage patterns correspond to the various modes of the outage distribution. The disconnection of the North-West occurs due to the non-local effects of cascading failures in areas not directly affected by high wind speeds. For example, hurricanes Harvey and Hanna never reach this region, but cause a considerable probability of outages affected by Harvey (Fig. 3(a)-(c)), but also due to non-local cascades as seen for Hanna (Fig. 3(h) and (i)). As the most populous city in Texas and a major load centre, the disconnection of Houston from the electrical networks causes the disconnection of a huge number of consumers from the electrical network and, consequently, the overproduction of generators located in the west of Texas, which have key roles to provide the required energy in Houston (see Supplementary Methods 5 and Supplementary Fig. 8). Interestingly, the northern part of the electric grid is never impacted by outages caused by these three hurricanes. Same figures for other hurricanes have been shown in Supplementary Fig. 7.

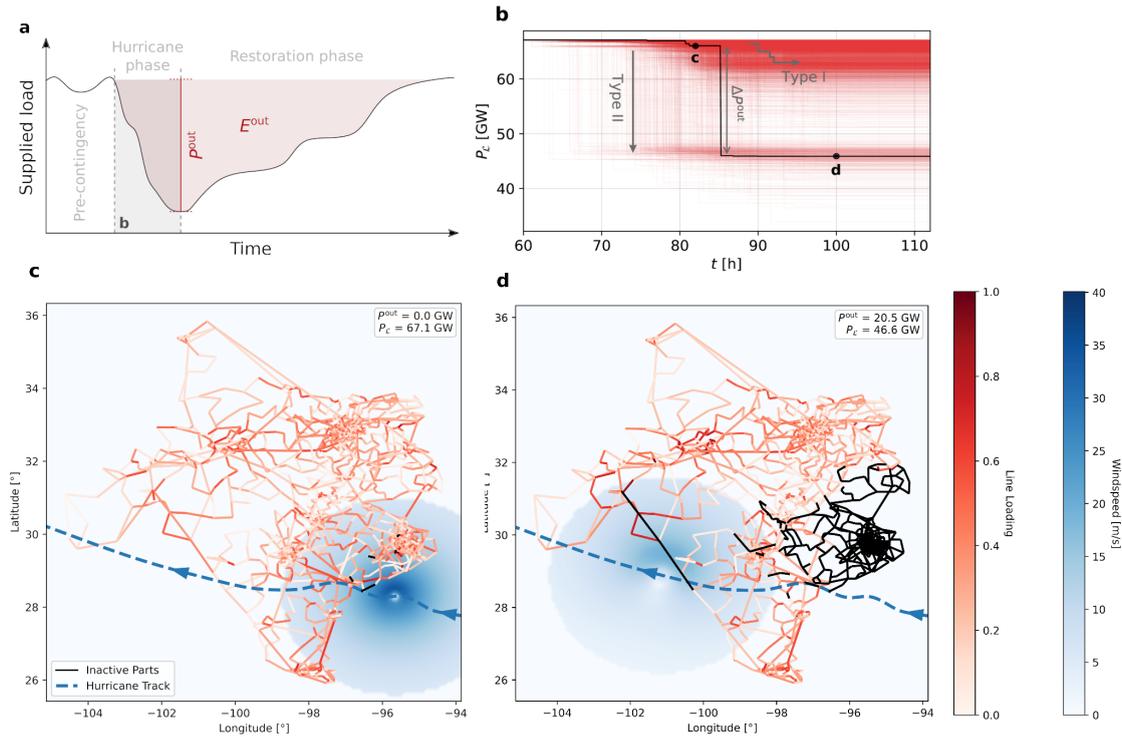

**Figure 2: Simulation of hurricane-induced cascading failures in the Texan electric grid** (a) The schematic variation of the supplied load in an electric grid before (pre-hurricane), during (hurricane phase) and after (restoration phase) a hurricane is loosely based on ERCOT's[23]. The total power outage $P^{out}$ after a hurricane has passed, and the total energy $E^{out}$ (red area) that was not supplied are measures for the severity of an outage scenario. (b) Summary of all realisations of power outage trajectories simulated for hurricane Claudette (see Methods section and Supplementary Methods 5 for specification of the model parameters). Trajectories shown in red come in two types, those that aggregate damages gradually over time (Type I in the figure) and those that include a large cascade (Type II). The distribution of cascade sizes is multimodal and we use an empirical threshold of $\Delta P^{out} > 15\,GW$ to define large cascades (see Supplementary Fig. 9). (c) and (d) show respectively the state of the power grid at the beginning and the end of the hurricane. These two states are shown in panel (b), for one realisation of primary line failures. Lines shown in black were destroyed by the hurricane or deactivated due to the secondary effects, for the other lines the relative line loading is shown, with red lines close to overload. In addition, the panel includes the track and a snapshot of the windfields of hurricane Claudette in blue. In the Supplement, we also provide a video of the simulation showing how the wind damages spread along the passage of hurricane Claudette.

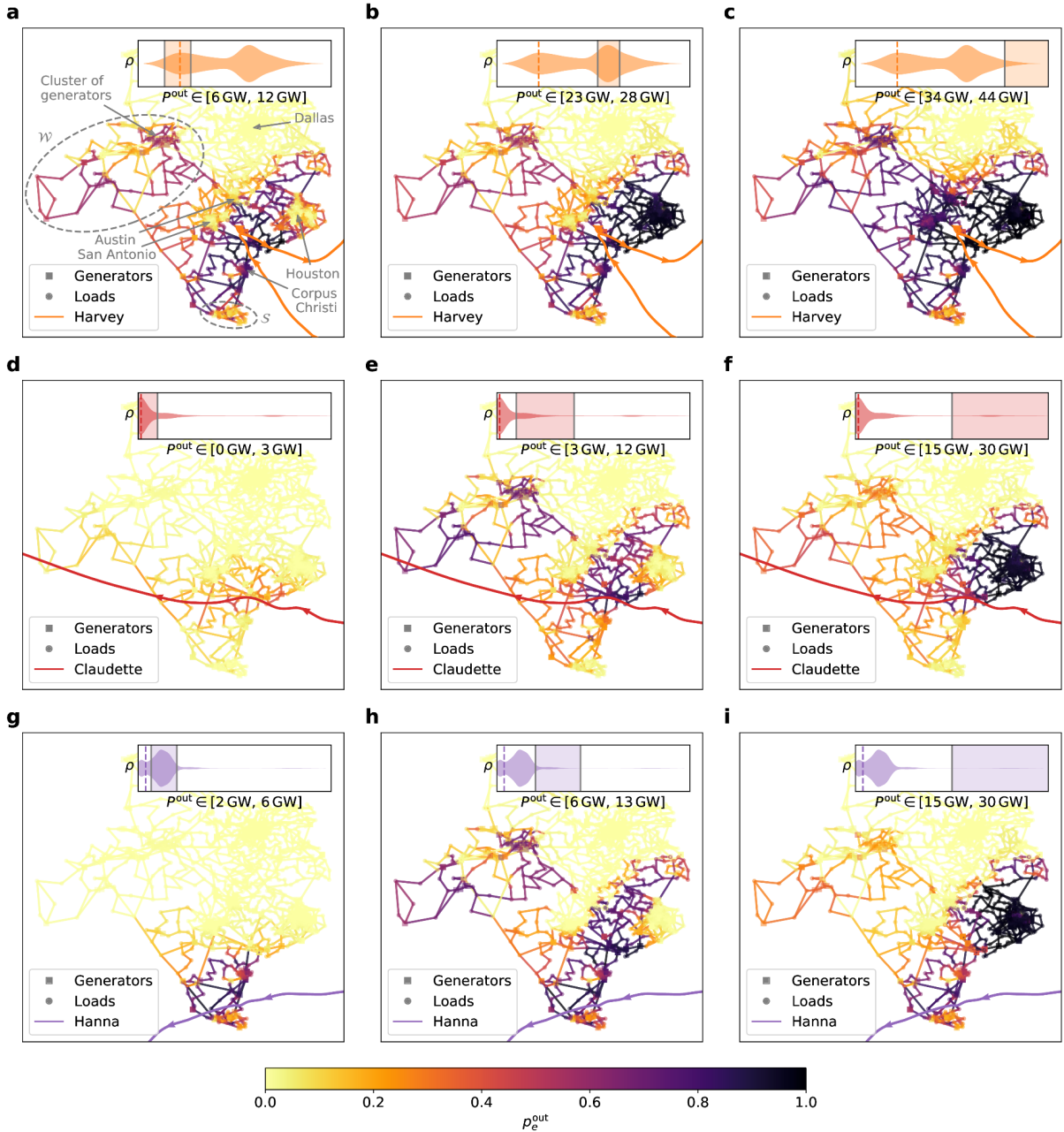

**Figure 4: Probability of line failure for different parts of the total power outage distribution** (a-i) Probability that the failure of a given power line is involved in three different modes of the power outage distribution. The modes are indicated by the insets and the exact range of considered power outages are shown below these insets. The probabilities $P_e^{out}$ are calculated as: number of realisations with a total outage within the specified range in each figure where the considered line failed / number of total realisations. The rows describe the probabilities for different hurricanes as indicated in the panel. Texan electric grid with grid elements colored according to their respective outage probability. The probability distributions shown in the insets are identical to the ones shown in Fig. 1(b).

For all seven hurricanes, the cascades play a major role in the total line failures associated with the event (see Fig. 4 ). They are induced by the overload of remaining lines and the isolation of grid elements, as well as the failure of islands with unavoidable overproduction.

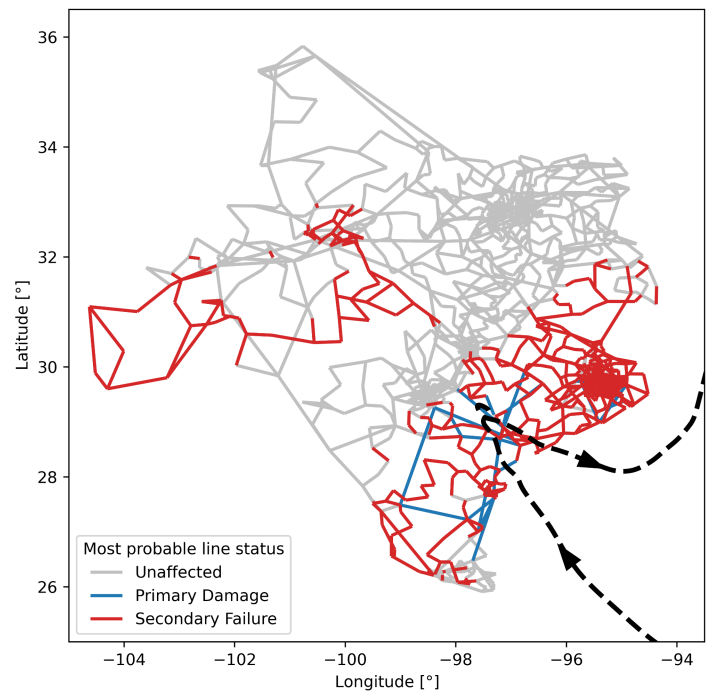

**Figure 3: The probability of primary damages and secondary failures induced by hurricane Harvey** In this plot the transmission lines are colored according to their high probability to be directly damaged by Harvey (blue lines) or to be deactivated due to the secondary effect of the hurricane (red lines). As expected the primary damages are located around the path of Harvey. However, secondary failures can occur far away from the hurricane track, which is related to the non-local effect of the primary damages in the power grid (see supplementary Methods 5). In this plot, grey lines have a higher probability of remaining operational than failing due to any reason.

Our results are not sensitive to the assumption of a homogeneous base failure rate as similar characteristics are also derived when assuming randomised base failure rates (see Supplementary Table 3). In addition, a temporal resolution of 5 minutes turned out to be adequate as time steps where several lines fail are rare. At this resolution it is also reasonable to assume that cascades of secondary failures have run their course before further lines are destroyed by the hurricane[24,25] (for further discussion regarding the temporal resolution, see Supplementary Note 2).

## Increasing Resilience

The fact that large cascades are triggered by the failure of specific lines suggests targeting these lines for protection. To identify the critical lines that should be protected we define a priority index as the probability that the wind-induced damage of this specific line triggers a large cascade, that is, a cascade that increase the outage by more than 15 GW, averaged over all seven hurricanes (see Fig. 2(b) and Eq. (4) in Methods).

As a baseline we also consider a conventional, static model (see Methods). The static index of a line is the conditional probability of a large outage given that the line is damaged by a TC. In both the co-evolution model and the static baseline (see Fig. 5(a) and (b)) the critical lines are mostly located around Houston.

To estimate the reduction in power outages that can be reached by protecting critical lines, we order them according to their priority index and evaluate the impact of the TC on the system with the first one to twenty lines protected, e.g. by being replaced by underground cables. It is worth noting that the co-evolution priority index value for most transmission lines is zero. Only $8\%$ of them have a value above $10^{-4}$, and only $20$ lines above $10^{-3}$. By protecting these $20$ lines, large power outages and cascading failures are almost completely prevented for smaller storms and dramatically reduced for the larger ones (see Fig. 5 and Supplementary Fig.9). For the stronger hurricanes Harvey and Ike, the power outage distributions are shifted from the second peak to the first peak with $P^{out} \leqslant 10\, GW$ (see Fig. 5(c)). Protecting the lines one by one shows that the reduction of the largest power outages improves smoothly, thus it is effective to protect up to twenty lines (see Fig. 5(c) and (d)). While in the original system damage amplification was almost guaranteed, it rarely occurs in the reinforced one. In summary $1\%$ of total lines reinforced leads to a 5 to 20 time reduction of the largest scale outages. The level of protection that can be reached by protecting the lines according to the priority index derived from the co-evolution models is generally higher than the protection of the same number of lines selected according to the priority index derived by the static model (see panel (d) of Fig. 5). The static baseline also identifies some of the most critical lines (see Supplementary 4), but additional protections stop being effective after the first 6-10 lines (see Fig. 5(c) and (d)). This demonstrates that the co-evolution model, with its detailed picture of the partially destroyed states, reveals genuinely new and critical information for increasing the resilience of the system.

It is worth to mention that the results obtained from homogeneous base failure rates are similar to the randomised ones (see Supplementary Methods 5 and Supplementary Table 3).

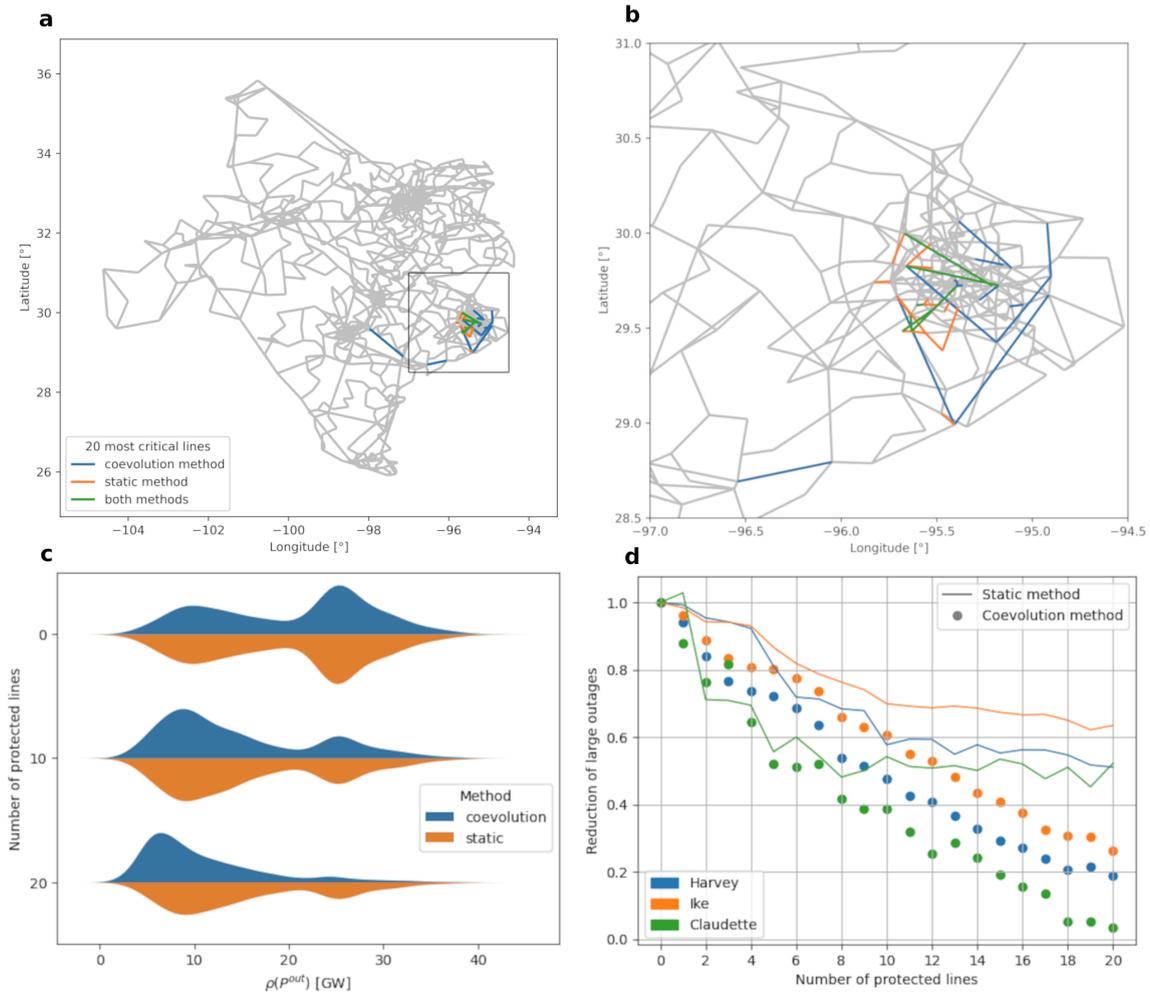

**Figure 5: Level of risk reduction that can be reached by protecting power lines according to the priority index: The co-evolution model against the static model** (a)-(b) 20 lines of the Texan power grid with the highest priority index (see Eq. [Eq. (4)](#)) obtained from the static model (orange lines), the co-evolution model (blue lines), and both approaches (green lines). The inset (b) shows a close-up view of Houston and Harris County, which contain most of the critical lines. As seen in (b) the critical lines obtained from both models are located in the same region, however, the co-evolutionary model identifies additional lines whose protection has a dramatic effect on increasing resilience. (c) Power outage distributions of hurricane Harvey in terms of the number of critical lines protected in both the co-evolution (blue) and the static model (orange). The second peak in the power outage distribution is strongly reduced as the number of protected lines increases. However, protecting lines obtained from the static model does not increase the resilience of the power grid as much as occurs in the co-evolution model. (d) Reduction of the large power outages obtained from both models. For all three strong hurricanes, i.e. Harvey, Ike and Claudette, the reduction in power outages is much greater in co-evolution model than the static one.

## Conclusion and outlook

The co-evolution model of the Texan power grid has been introduced as an efficient approach to temporally resolve the line failures and secondary grid outages induced by TCs. The model can resolve to considerable detail the way secondary failure cascades amplify the impact of extreme events. Using this information it can be used to identify critical lines that should be protected to effectively increase the system's resilience and prevent the most severe outages. Our model goes significantly beyond the state of the art so far represented by statistical and economic models that can only capture a static picture of the event and the network[24–29]. We have seen that such static approaches do not easily identify all of the critical lines during extended events. Their importance is only revealed by stepwise 'tracking' the destruction of the system and associated power outages and overloads. We expect that this co-evolution approach will also be a promising tool to understand and protect other grids exposed to spatio-temporally extended extreme events.

The results of our study are in agreement with a recent TC related risk assessment for Texas[26]. Combining our priority index with additional information about the cost of a reinforcement of the considered lines could also enable the identification of the most cost efficient way to reduce the probability of power outages above a critical limit to an intended value (see Supplementary Methods 6).

While the model based on wind speeds and historical hurricane tracks already identified crucial structures in the grid, the co-evolution approach could naturally be extended to more sophisticated models and broader settings. One particularly important goal for future research will be to drive the model with potential future storm tracks due to climate change[27]. As the frequency of particularly strong TCs is expected to increase under global warming (WGI contribution to the AR6), understanding what lines are critical in the face of the weather of the next decades is crucial. Another important avenue of broadening the model is to account for TC induced flooding (coastal flooding, pluvial or fluvial flooding) and associated destructions. These may follow a different temporal pattern where the adequacy of the approach proposed here has to be newly tested. This would also provide a first step towards an assessment of genuine compound events in which several stresses for the grid coincide.

# Methods

## Electric grid data of Texas

For the study we used the publicly available electric grid test case *ACTIVSg2000*[28], that covers the area of the so-called ERCOT Interconnection, which supplies $90$ percent of the electricity demand in Texas[29]. The test case is synthetic but resembles fundamental properties of the real grid, such as the spatial distribution of power generation and demand[21]. It encompasses $2000$ buses with geographic locations, $3116$ branches (both transmission lines and transformers) and covers four different voltage levels. The test case comes with all required electrical parameters ranging from the power injections of buses to the power flow capacities of transmission lines and transformers. The flow capacities $C_{ij}$ play a particularly important role for the simulation of cascading failures as they determine the amount of power that can be transported by individual lines and transformers without potentially damaging the equipment.

## Historical hurricane data

Hurricane storm tracks are extracted from the International Best Track Archive for Climate Stewardship (IBTrACS)[30, 31] as time series of cyclone center coordinates along with meteorological variables like maximum sustained wind speeds and minimum pressure on a $3-6$ h snapshot basis. For this study, a hand-picked selection of seven historical storms is used (Supplementary Fig. 1 and 2) to cover several different types of trajectories and intensities. Particularly, the selection also includes storms that continue to move westward after landfall and affect the southern and western parts of Texas (see Hurricane Claudette, Tropical Storm Erin, and Hurricane Hanna in Supplementary Fig. 2 and the Supplementary Fig. 1), contrary to most hurricanes that are steered northward by the Coriolis effect before western parts of Texas are reached[22]. From the track records, we compute time series of wind fields within a radius of $300$ km from the storm center using the Holland model for surface winds, as implemented in the Python-package CLIMADA[32, 33], at a spatial resolution of $0.1$ degrees (approximately $11$ km) and a temporal resolution of $5$ minutes. The intensities of the considered storms are also shown along the respective tracks in Supplementary Fig.1 while other properties of the storms are listed in Supplementary Table 1.

## Transmission line fragility model

To model wind-induced failures of transmission lines, we first differentiate between overhead transmission lines and underground cables in the electric grid of Texas. Following Birchfield et al., we analyse lines that are shorter than $12.875$ km ($8$ miles) and connect a total load of at least $200$ MW as underground cables[21]. All other lines are assumed to be overhead transmission lines. The latter are then divided into segments of length $l \approx 161$ m, which corresponds to the average distance between transmission towers in Texas[34]. Our fragility model assigns failure rates to individual line segments $k$ according to

$$r_k(v_t) = r^{\text{brk}} \frac{F_k^{\text{wind}}(v_t)}{F^{\text{brk}}}, (1)$$

where, $F_k^{\text{wind}}$ denotes the wind force acting on the line segment $k$ for a given wind speed $v_t$ and is calculated according to the guidelines published by the American Society of Civil Engineers[35]. The parameter $r^{\text{brk}}$ represents the inverse of the so-called time to failure, which indicates how long a line segment can withstand a wind force equal to the breaking force $F^{\text{brk}}$. It is used as a free parameter to calibrate the model such that historically reported power outages are reproduced in our simulations (see Supplementary Methods 5). The full wind force equation as well as the meaning and the values of all parameters can be found in Supplementary Methods 2 and Supplementary Table 2. In all figures shown in the main text, $r^{brk} = 0.002\ h^{-1}$. Using the failure rates $r_k$, we define the probability that a line segment $k$ fails during the time interval $[t, t+\tau)$ as

$$p_k(v_t) = \min\left(r_k(v_t) \cdot \tau,\ 1\right). \quad (2)$$

This failure probability is inspired by the line fragility model established by Winkler et al., which assumes that the failure probability is proportional to the ratio of the wind force and the breaking force[36]. However, in contrast to their model, we define the failure probability $p_k$ using a time-dependent failure rate $r_k$ that allows us to take the time evolution of a field into account. A line is removed from the test case if any of its line segments fails during a time interval. It should be noted that multiple lines may be destroyed in the same time step, meaning that they are removed from the network simultaneously. According to Eq. (2), the probability of simultaneous failures increases with time step size $\tau$. A discussion of the role of the time resolution can be found in Supplementary Note 2.

## Cascading failure model

Wind-induced line failures can trigger cascades of overload failures in the branches of the electric grid. As cascading failures typically evolve on smaller time scales than the temporal resolution $\tau$ of the wind field, we can assume a time scale separation. When the network topology is changed by a primary damage event, the power flows $P_{ij}$ on the branches are rerouted using the DC power flow model

$$P_i = \sum_{j=1} P_{ij} = \sum_{j=1}^{N} B_{ij}(\theta_i - \theta_j), \quad (3)$$

here, $P_i$ are the net active power injections at the buses, $\theta_i$ are the bus voltage angles and $B_{ij}$ are the elements of the nodal susceptance matrix that comprises the network topology. More details on the assumptions of the DC power flow model and the software used can be found in Supplementary Methods 3. If the new state of the network exhibits any overloaded branch ($|P_{ij}| > C_{ij}$), they are deactivated and the process is repeated. When the network reaches a state without overloads, the algorithm advances to the

next primary damage event. When a load or generator gets disconnected or the grid is split into several parts, the global active power balance (GAPB) has to be restored in each network component. Motivated by a primary frequency control in real electric grids, we adjust the outputs of generators uniformly, while respecting their output limits defined in the data set. Whenever the generator limits do not allow to fully restore the GAPB, we either conduct a uniform minimal load shedding or consider the blackout of the whole network component in the case of an unavoidable overproduction. The details of the algorithm are explained in Supplementary Methods 4.

## Quantification of power outages

We use the following three different quantities to track the power outages arising in our simulations: (i) $P_{\mathcal{L}}(t)$ denotes the total supplied load at the end of each time step, i.e., after the cascading algorithm finished, respectively. It is calculated by adding up the demands of all connected loads across all islands that exist at the given time. Since our co-evolution model assumes that cascading failures happen instantaneously, $P_{\mathcal{L}}(t)$ represents a step function for each individual TC scenario as shown in Fig. 2(b). We have simulated $10^4$ scenarios for each hurricane. (ii) Any cascading failure that actually causes a loss of supplied load results in a vertical transition of size $\Delta P^{\text{out}}$ in $P_{\mathcal{L}}(t)$. One such transition is annotated with $\Delta P^{\text{out}}$ for the highlighted scenario in Fig. 2(b). (iii) All cascading failures that are triggered in a given TC scenario lead to a final power outage $P^{\text{out}} = P_{\mathcal{L}}^{\text{init}} - P_{\mathcal{L}}^{\text{final}} \in [0\,GW, 67.1\,GW]$. The interesting statistics of $P^{\text{out}}$ are shown and discussed in Fig. 1(b).

## Identification of critical lines

We identify critical overhead transmission lines by means of a priority index defined for each line $(i,j)$ as

$$\kappa_{ij} := \frac{1}{|\mathcal{H}|} \sum_{h \in \mathcal{H}} p_{ij}^{(\text{II})}(h), \quad (4)$$

where $\mathcal{H}$ denotes the set of considered hurricanes (seven hurricanes in this study) and $p_{ij}^{(\text{II})}$ is the probability of a large cascade being triggered by the wind-induced failure of line $(i,j)$. More specifically, we call cascades large or belonging to type II if their associated power outage $\Delta P^{\text{out}}$ lies above an empirical threshold of $15$ GW (indicated as type II in Fig. 2(b) and Fig. 5(d)). Eq. (4) includes an averaging over all considered hurricanes to discern lines that are critical for multiple hurricanes. This allows us to propose line reinforcements that increase the resilience not only for a particular hurricane. Some properties of the $20$ most critical lines found in this study are listed in

Supplementary Table 3. [Fig. 5(a) and (b)](#) shows the location of these lines and demonstrates that reinforcing them indeed increases the resilience of the electric grid substantially. More details of the critical lines and a possibility to incorporate economic considerations into our analysis are discussed in Supplementary Methods 6.

## Baseline Method

Here, we apply the static model as a baseline method. By static model, we mean that all primary damages occur simultaneously and, then, the DC power model along with global active power balance (see Supplementary Methods 6) are activated once to bring back the energy balance in the system and to evaluate the total final power outages $P^{out}$. As discussed in Supplementary Note 2 the final power outage distributions are independent of the time resolution of the wind field, however the primary damages leading to large outages, i.e. $20\,GW$ to $30\,GW$, can be completely different ones. To indicate the critical lines obtained from the static model, first, we separate all scenarios in which $P^{out} > 15\,GW$. Then, we use [Eq. (4)](#) to calculate the priority index of the primary damages leading to large cascades. The top $20$ lines with the highest priority index have been listed in Supplementary Table 5. As seen in this table, except for the six lines highlighted in red, the other lines are completely different from lines obtained from the co-evolution model.

## Code availability

All code necessary to reproduce the findings in this work is openly available. The time-dependent wind fields are computed using the open-source platform CLIMADA[32, 33]. The implementation of the transmission line fragility and the DC power model is available from https://gitlab.pik-potsdam.de/stuermer/itcpg.jl.

## Data availability

The observed TCs from IBTrACS[30, 31] are distributed under the permissive WMO open data licence through the IBTrACS website (https://www.ncei.noaa.gov/products/international-best-track-archive) and can be directly retrieved through the CLIMADA[32, 33] platform. The electrical network data is openly available from the Texas A&M University's electric grid test case repository (https://electricgrids.engr.tamu.edu/electricgrid-test-cases/activsg2000/).

## Acknowledgements

This project has received funding from the ConNDyNet2 project under grant no. 03EF3055F. This research has received funding from the German Academic Scholarship Foundation and the German Federal Ministry of Education and Research (BMBF) under the research projects QUIDIC (01LP1907A) and SLICE (FKZ: 01LA1829A), and from the CHIPS project, part of AXIS, an ERA-NET initiated by JPI Climate, funded by FORMAS (Sweden), DLR/BMBF (Germany, grant no. 01LS1904A), AEI (Spain) and ANR (France) with co-funding by the European Union (grant no. 776608).

## Author Contribution

M. Anvari, F. Hellmann and C. Otto contributed to design and conceive the research. The co-evolution model is designed and developed by M. Anvari, J. Stürmer, A. Plietzsch and F. Hellmann. All simulations and data analyses of this work have been done by J. St☐ürmer and under supervision of M. Anvari. All hurricane data have been provided by T. Vogt during this research. All authors contributed to discussing and interpreting the results, and contributed to writing the manuscript.

## Competing Interests

The authors declare that they have no competing interests.